\documentclass[]{IAC_style}

\usepackage{bm}
\usepackage{commath}
\usepackage{stfloats}

\usepackage{amsmath,amsfonts}
\usepackage{algorithmic}
\usepackage{algorithm}
\usepackage{array}
\usepackage[caption=false,font=normalsize,labelfont=sf,textfont=sf]{subfig}
\usepackage{textcomp}
\usepackage{stfloats}
\usepackage{url}
\usepackage{verbatim}
\usepackage{graphicx}
\usepackage{cite}
\usepackage{subcaption}
\usepackage{comment}
\usepackage[super]{nth}
\usepackage[version=4]{mhchem}
\usepackage[utf8]{inputenc}
\usepackage{siunitx}
\usepackage{amsmath,amsthm,amssymb}
\usepackage{longtable,tabularx}
\usepackage{bm}
\usepackage{commath}
\usepackage{hyperref}
\usepackage{tikz}
\usetikzlibrary{arrows.meta,calc}

\newtheorem{definition}{Definition}[section]

\usepackage{lipsum}

\begin{document}


\IACconference{77}
\IACpaperyear{2026}
\IACpapernumber{IAC-26,C1,LBA,3,x117776}
\IAClocation{Antalya, T\"urkiye}
\IACdate{5-9 October 2026}
\IACcopyrightA{}

\title{Fuel-Efficient Control Barrier Function-Embedded Model Predictive Control for Autonomous Spacecraft Rendezvous and Docking}

\IACauthor{Sabire Betül Çınar}{1}{0}
\IACauthor{Ege Ç. Altunkaya$^{\orcidlink{0000-0001-8938-7632}}$}{2}{0}
\IACauthor{Abdülbaki Şanlan}{1}{0}
\IACauthor{Emre Koyuncu$^{\orcidlink{0000-0002-0726-4979}}$}{1}{0}
\IACauthor{İbrahim Özkol$^{\orcidlink{0000-0002-9300-9092}}$}{1}{0}

\IACauthoraffiliation{
Faculty of Aeronautics and Astronautics,
Istanbul Technical University,
Istanbul, Türkiye, 34469.
}

\IACauthoraffiliation{
Aviation Institute,
Istanbul Technical University,
Istanbul, Türkiye, 34469.
}

\abstract{

Autonomous spacecraft rendezvous and docking with a rotating target requires the chaser to track a moving docking port while satisfying time-varying safety and actuator constraints. The problem is further complicated by uncertainty in the target attitude and angular rate, which directly affects the line-of-sight (LoS) corridor. This work proposes a robust control barrier function-embedded model predictive control framework for planar rendezvous and docking. Relative motion is modeled in the LVLH frame, while a backstepping-based tracking objective regulates position and velocity with respect to the moving docking port. The rotating LoS corridor is enforced over the MPC prediction horizon using relative-degree-two exponential control barrier functions. Target-orientation uncertainty is handled by contracting the admissible LoS corridor according to the propagated uncertainty bound, without altering the nominal prediction model. Closed-loop simulations with nonlinear relative dynamics show that the controller maintains LoS safety, respects thrust limits, and drives the chaser toward the desired docking condition despite increasing orientation uncertainty.

}

\maketitle
\thispagestyle{fancy}


\section{Introduction}
\label{introduction}

Autonomous rendezvous and docking is an enabling capability for missions such
as on-orbit servicing, debris removal, spacecraft recovery, and in-space
assembly. The terminal approach is particularly demanding, since the chaser
must accurately converge to the docking interface while simultaneously
satisfying geometric and actuation constraints. This challenge becomes more
pronounced when the target is rotating and its motion is not known exactly.

\subsection{Related Work}

Previous studies have addressed autonomous spacecraft rendezvous and docking
through predictive and safety-oriented control approaches. MPC has been widely
adopted because it allows state and input constraints to be treated directly
within a receding-horizon optimization problem. In particular, it has been
used to enforce thrust, line-of-sight (LoS), soft-docking, and
collision-avoidance constraints, while also accounting for the evolution of
rotating docking geometries and obstacle configurations
\cite{park2011mpc,bashnick2023fastmpc,fear2024mpc}. Building on this
predictive framework, robust MPC formulations have been developed for
non-cooperative and tumbling targets, where uncertainty in the target state,
motion, or model parameters is addressed through tube-based prediction and
integrated estimation--planning--control architectures
\cite{specht2023tubempc,albee2021robust}. Safe MPC formulations have further
extended this line of work by addressing recursive feasibility together with
explicitly time-varying LoS and collision-avoidance constraints for docking
with tumbling targets \cite{dong2022safempc}.

In parallel, complementary approaches have focused more directly on safety and
efficiency guarantees. Input-constrained CBF formulations provide a means of
identifying control-invariant safe subsets and enforcing safety despite bounded
actuator authority \cite{agrawal2021iccbf}. Barrier-based safety enforcement
has also been demonstrated in aerospace applications, where adaptive
exponential CBFs have been combined with flight-envelope protection to enforce
collision-avoidance and dynamic-limit constraints under bounded control
authority \cite{altunkaya2025nuisance}. Stochastic optimization approaches have
examined the trade-off between LoS safety and fuel expenditure under modeling
and sensing uncertainty \cite{vinod2018tradeoff}. Real-time collision-free
guidance has also been investigated using adaptive artificial potential fields,
with emphasis on low computational burden and onboard applicability
\cite{zappulla2019aapf}, while fuel-efficient terminal docking to uncertain
tumbling targets has been studied through robust trajectory-generation methods
\cite{sternberg2017docking}. Taken together, these studies motivate a framework
that combines the predictive constraint-handling capability of MPC with
explicit barrier-based LoS safety under target-motion uncertainty.

\subsection{Problem Statement \& Objectives}

\begin{figure}[htb!]
    \centering
    \includegraphics[width=\columnwidth]{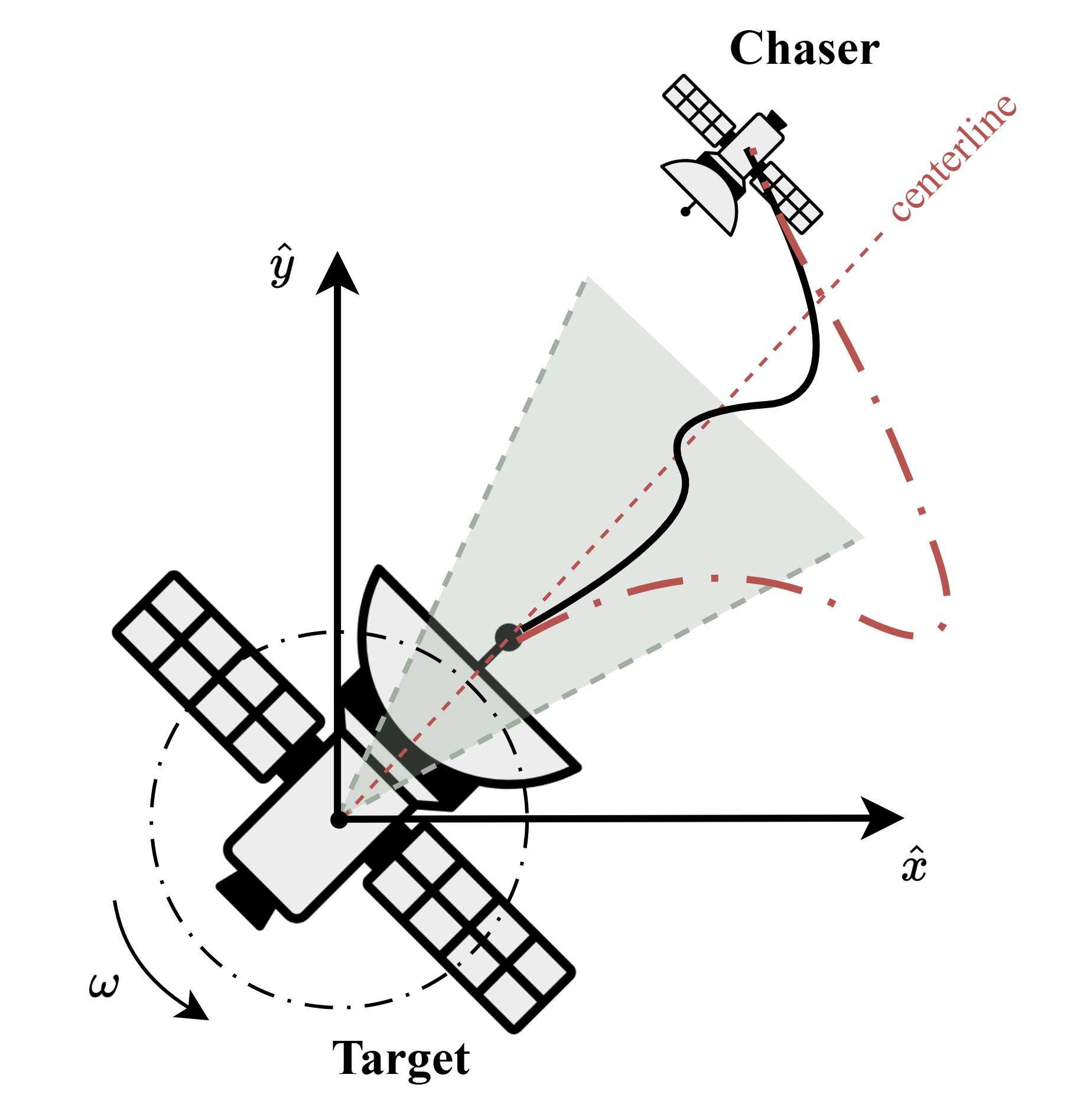}
    \caption{Planar rendezvous and docking problem considered in this study.
    The chaser spacecraft approaches the rotating target and tracks the moving
    docking port while remaining inside the prescribed LoS corridor. The
    rotation of the target causes both the docking geometry and the associated
    safety corridor to evolve throughout the maneuver.}
    \label{fig:problemStatement}
\end{figure}

Consider the planar rendezvous and docking problem illustrated in
Fig.~\ref{fig:problemStatement}, where a chaser spacecraft approaches a target
spacecraft and subsequently tracks a docking port located on the target body.
The target is subject to a planar tumbling motion, represented in this study
by a constant angular rate relative to the LVLH frame. Consequently, the
docking port and the associated line-of-sight (LoS) corridor rotate
continuously throughout the maneuver.

The LoS requirement is treated as a safety constraint rather than as an
additional tracking objective. Since the target orientation is propagated
forward from imperfect attitude and angular-rate information, the uncertainty
associated with the corridor orientation grows over the prediction interval.
A predictive control architecture is therefore employed so that the future
motion of the docking reference, the rotation of the safety corridor, and the
corresponding uncertainty growth can be considered before the control action
is applied. In particular, the LoS requirement is formulated through control
barrier functions and imposed along the MPC prediction horizon, thereby
allowing the available control authority to be allocated while anticipating
future safety restrictions rather than reacting solely to the instantaneous
constraint.

Accordingly, the control objective is to determine a bounded thrust command
such that the chaser approaches the moving docking port while remaining inside
the uncertainty-consistent LoS corridor. In terms of the tracking errors
introduced later, the desired closed-loop behavior is
$\bm e_r(t)\rightarrow\bm 0$,
$\bm e_v(t)\rightarrow\bm 0$,
$\bm X(t)\in C_{\mathrm{LoS}}^r(t)$, and
$\|\bm u(t)\|_\infty\leq u_{\max}$.
The present study therefore focuses on robustifying the safety constraint
against the propagated orientation uncertainty; robustification of the
docking-reference tracking problem itself is outside the considered scope.

\subsection{Contributions \& Organization}

The main contributions of this study are summarized as follows:

\begin{itemize}

    \item A CBF-embedded MPC architecture is formulated for planar rendezvous
    and docking with a rotating target, where a backstepping-based tracking
    objective regulates both position and velocity relative to the moving
    docking port while the rotating relative-degree-two LoS constraints are
    enforced over the prediction horizon.

    \item A geometry-based robust LoS formulation is developed for bounded
    target-attitude and angular-rate knowledge errors. The propagated
    uncertainty is incorporated through a time-varying contraction of the
    admissible corridor, thereby confining the robustification to the safety
    constraint without modifying the nominal prediction dynamics or tracking
    objective.

\end{itemize}

The remainder of the paper is organized as follows.
Section~\ref{preliminaries} introduces the mathematical preliminaries and the
relative-motion model. Section~\ref{methodology} presents the
backstepping-based MPC tracking formulation, the rotating LoS barrier
construction, and the uncertainty-aware safety constraint.
Section~\ref{results} evaluates the proposed architecture in the nonlinear
rendezvous and docking simulation, followed by the concluding remarks and
future research directions.

\section{Preliminaries}
\label{preliminaries}

\subsection{Notations}

A vector is denoted in bold type, i.e., $\bm v$, and the time derivative of a
function is denoted by $\dot{\{\}}$. The transpose and pseudo-inverse operators
for a matrix are denoted by $\{\}^{\top}$ and $\{\}^{\dagger}$, respectively,
whereas $\bm I_n$, $\bm 0_n$, and $\bm 0_{n\times m}$ denote the identity
matrix, zero vector, and zero matrix of appropriate dimensions. The Euclidean
and $\ell_\infty$ norms are denoted by $\|\{\}\|_2$ and $\|\{\}\|_\infty$,
respectively. For a symmetric positive-definite matrix $\bm M$, the weighted
quadratic norm is denoted by
$\|\bm x\|_{\bm M}^2 \triangleq \bm x^\top \bm M \bm x$.
The notation $\mathcal{K}_\infty^e$ denotes the set of extended
class-$\mathcal{K}_\infty$ functions.

Consider the control-affine system
\begin{equation}
\label{controlAffineSystem}
    \dot{\bm x}
    =
    \bm f(\bm x)
    +
    \bm g(\bm x)\bm u,
\end{equation}
where $\bm x\in\mathbb{R}^n$ and
$\bm u\in U\subset\mathbb{R}^m$ denote the state and admissible control input,
respectively, and
$\bm f:\mathbb{R}^n\rightarrow\mathbb{R}^n$ and
$\bm g:\mathbb{R}^n\rightarrow\mathbb{R}^{n\times m}$ are locally Lipschitz
continuous. Finally, consider a control-affine strict-feedback system with
state
$(\bm{x},\bm{\xi})\in\mathbb{R}^{n}\times\mathbb{R}^p$

\begin{equation}
\begin{split}
\label{strictFeedback}
    \dot{\bm x}
    &=
    \bm f_0(\bm x)
    +
    \bm g_0(\bm x)\bm \xi,
    \\
    \dot{\bm \xi}
    &=
    \bm f_1(\bm x,\bm \xi)
    +
    \bm g_1(\bm x,\bm \xi)\bm u,
\end{split}
\end{equation}
where $\bm \xi\in\mathbb{R}^p$, and nonlinear mappings
$\bm f_0:\mathbb{R}^n\to\mathbb{R}^n$,
$\bm g_0:\mathbb{R}^n\to\mathbb{R}^{n\times p}$,
$\bm f_1:\mathbb{R}^n\times\mathbb{R}^p\to\mathbb{R}^p$, and
$\bm g_1:\mathbb{R}^n\times\mathbb{R}^p\to\mathbb{R}^{p\times m}$
are locally Lipschitz continuous. Moreover, assume $p\leq m$ and
$\mathrm{rank}\big(\bm g_1(\bm x,\bm \xi)\big)=p$ on the domain of interest so
that $\bm g_1(\bm x,\bm \xi)$ admits a right inverse; in particular,
$\bm g_1^{-1}$ is used when $p=m$ and $\bm g_1$ is nonsingular, whereas the
Moore--Penrose pseudo-inverse $\bm g_1^\dagger$ is used when $p<m$.

\subsection{Local-Vertical Local-Horizontal Relative Dynamics}
\label{lvlhDynamicsSection}

Consider a chaser spacecraft of mass $m_c$ performing planar relative motion
with respect to a target spacecraft moving on a circular orbit of radius
$r_0$. The orbital mean motion is $n=\sqrt{\mu/r_0^3}$, where $\mu$ is the
gravitational parameter of the central body. Let
$\bm r=[x,y]^\top$ and
$\bm v=[v_x,v_y]^\top$ denote the chaser position and velocity relative to the
target in the local-vertical local-horizontal (LVLH) frame, and define
$\bm X=[\bm r^\top,\bm v^\top]^\top\in\mathbb{R}^4$.
The commanded thrust is denoted by $\bm u=[u_x,u_y]^\top$.

The exact planar nonlinear relative dynamics are
\begin{equation}
\label{nonlinearLVLH}
\begin{split}
    \ddot{x}
    &=
    n^2x+2n\dot y
    +\frac{\mu}{r_0^2}
    -\frac{\mu(r_0+x)}
    {\big((r_0+x)^2+y^2\big)^{3/2}}
    +\frac{u_x}{m_c},
    \\
    \ddot{y}
    &=
    n^2y-2n\dot x
    -\frac{\mu y}
    {\big((r_0+x)^2+y^2\big)^{3/2}}
    +\frac{u_y}{m_c}.
\end{split}
\end{equation}

For $\|\bm r\|_2\ll r_0$, linearization of~\eqref{nonlinearLVLH} about the
target circular orbit yields the planar Clohessy--Wiltshire (CW) model
\begin{equation}
\label{lvlhDynamics}
    \dot{\bm X}
    =
    \underbrace{
    \begin{bmatrix}
        0 & 0 & 1 & 0\\
        0 & 0 & 0 & 1\\
        3n^2 & 0 & 0 & 2n\\
        0 & 0 & -2n & 0
    \end{bmatrix}}_{\bm A}
    \bm X
    +
    \underbrace{
    \begin{bmatrix}
        \bm 0_{2\times2}\\
        \dfrac{1}{m_c}\bm I_2
    \end{bmatrix}}_{\bm B}
    \bm u.
\end{equation}

The nonlinear dynamics in~\eqref{nonlinearLVLH} are employed for closed-loop
propagation, whereas~\eqref{lvlhDynamics} constitutes the prediction model of
the MPC. For subsequent derivations, the acceleration component of
~\eqref{lvlhDynamics} is written as
\begin{equation}
\label{lvlhPartitioned}
\begin{aligned}
    \dot{\bm v}
    &=
    \bm A_r\bm r
    +
    \bm A_v\bm v
    +
    \frac{1}{m_c}\bm u,
    \\
    \bm A_r
    &=
    \begin{bmatrix}
        3n^2&0\\
        0&0
    \end{bmatrix},
    \qquad
    \bm A_v
    =
    \begin{bmatrix}
        0&2n\\
        -2n&0
    \end{bmatrix}.
\end{aligned}
\end{equation}

The target is allowed to rotate relative to the LVLH frame at a constant
angular rate $\omega$. The docking-port bearing is therefore
$\psi(t)=\psi_0+\omega t$. Defining
$\bm e(\theta)=[\cos\theta,\sin\theta]^\top$ and
$\bm S\triangleq
\left[\begin{smallmatrix}0&-1\\1&0\end{smallmatrix}\right]$,
the docking-port kinematics are
\begin{equation}
\label{portKinematics}
\begin{aligned}
    \bm r_p(t)
    &=
    \rho_p\bm e\big(\psi(t)\big),
    &
    \bm v_p(t)
    &=
    \omega\bm S\bm r_p(t),
    \\
    \bm a_p(t)
    &=
    -\omega^2\bm r_p(t).
\end{aligned}
\end{equation}
where $\rho_p$ denotes the distance of the docking port from the target center.

\subsection{Lyapunov Backstepping}

Consider the control-affine strict-feedback system defined in
~\eqref{strictFeedback}. Define a continuously differentiable and positive
definite Lyapunov function
$V_0(\bm x):\mathbb{R}^n\to\mathbb{R}_{\geq0}$.
A designed stabilizing controller
$\bm k_0(\bm x):\mathbb{R}^n\to\mathbb{R}^p$ with
$\bm k_0(\bm0)=\bm0$ satisfies
\begin{equation}
\label{backsteppingGeneral1}
    \frac{\partial V_0}{\partial\bm x}(\bm x)
    \Big(
    \bm f_0(\bm x)
    +
    \bm g_0(\bm x)\bm k_0(\bm x)
    \Big)
    \leq
    -\gamma\big(\|\bm x\|_2^2\big),
\end{equation}
where $\gamma\in\mathcal K_\infty^e$. Since the state vector to be controlled
$\bm\xi$ is directly controlled by the physical input $\bm u$, one must apply
a further derivation to design the stabilizing controller
$\bm k(\bm x,\bm\xi)$. For this purpose, define the composite Lyapunov function
$V_1:\mathbb{R}^n\times\mathbb{R}^p\to\mathbb{R}_{\geq0}$ as
\begin{equation}
\label{backsteppingGeneral2}
    V_1(\bm x,\bm \xi)
    =
    V_0(\bm x)
    +
    \frac{1}{2}
    \big(\bm\xi-\bm k_0(\bm x)\big)^\top
    \big(\bm\xi-\bm k_0(\bm x)\big).
\end{equation}

There exists a stabilizing Lipschitz continuous controller
$\bm k(\bm x,\bm\xi)$ for all
$\bm x\in\mathbb{R}^n$ and $\bm\xi\in\mathbb{R}^p$ of the form
\begin{equation}
\label{backsteppingGeneral3}
\begin{aligned}
\bm k(\bm x,\bm\xi)
={}&
\bm g_1^\dagger(\bm x,\bm\xi)
\Bigg[
-\bm f_1(\bm x,\bm\xi)
+
\frac{\partial\bm k_0}{\partial\bm x}(\bm x)
\Big(
\bm f_0(\bm x)
+
\bm g_0(\bm x)\bm\xi
\Big)
\\
&\hspace{1.0cm}
-
\left(
\frac{\partial V_0}{\partial\bm x}(\bm x)
\bm g_0(\bm x)
\right)^\top
-
\lambda
\big(
\bm\xi-\bm k_0(\bm x)
\big)
\Bigg].
\end{aligned}
\end{equation}
This controller yields
\begin{equation}
\label{backsteppingGeneral4}
\begin{aligned}
    \dot V_1
    ={}&
    \frac{\partial V_0}{\partial\bm x}(\bm x)
    \Big(
    \bm f_0(\bm x)
    +
    \bm g_0(\bm x)\bm k_0(\bm x)
    \Big)
    \\
    &-
    \lambda
    \big(
    \bm\xi-\bm k_0(\bm x)
    \big)^\top
    \big(
    \bm\xi-\bm k_0(\bm x)
    \big)
    \\
    \leq{}&
    -\gamma\big(\|\bm x\|_2^2\big)
    -
    \lambda
    \|\bm\xi-\bm k_0(\bm x)\|_2^2
    \leq0,
\end{aligned}
\end{equation}
where $\lambda\in\mathbb{R}_{>0}$. Thereby, $V_1$ is a composite Lyapunov
function for~\eqref{strictFeedback}, and the equilibrium
$(\bm x,\bm\xi)=(\bm0_n,\bm0_p)$ is asymptotically stable; hence
$\bm x(t)\to\bm0_n$ and $\bm\xi(t)\to\bm0_p$ as $t\to\infty$ for all initial
conditions in its domain of attraction.

\subsection{Control Barrier Functions}

To encode state constraints, let
$h:D\subset\mathbb{R}^n\to\mathbb{R}$ be a continuously differentiable barrier
function defined on a domain $D$. The associated safe set is
\begin{equation}
    C
    \triangleq
    \left\{
    \bm x\in D
    \mid
    h(\bm x)\geq0
    \right\},
\end{equation}
with boundary
$\partial C\triangleq\{\bm x\in D\mid h(\bm x)=0\}$ and interior
$\mathrm{Int}(C)\triangleq\{\bm x\in D\mid h(\bm x)>0\}$.
The control objective is to render $C$ forward invariant.
Consider the control-affine system in~\eqref{controlAffineSystem} with
admissible inputs $\bm u\in U\subset\mathbb{R}^m$. The Lie derivatives of
$h$ along $\bm f$ and $\bm g$ are
\begin{equation}
\begin{split}
    \mathcal L_fh(\bm x)
    &\triangleq
    \nabla h(\bm x)^\top\bm f(\bm x),
    \\
    \mathcal L_gh(\bm x)
    &\triangleq
    \nabla h(\bm x)^\top\bm g(\bm x).
\end{split}
\end{equation}

\begin{definition}[Control Barrier Function \cite{ames2017cbf}]
A continuously differentiable function
$h:D\to\mathbb{R}$ is a control barrier function on $D$ if there exists an
extended class-$\mathcal K_\infty$ function $\alpha$ such that, for all
$\bm x\in D$,
\begin{equation}
\label{cbf1}
    \sup_{\bm u\in U}
    \left\{
    \mathcal L_fh(\bm x)
    +
    \mathcal L_gh(\bm x)\bm u
    \right\}
    \geq
    -\alpha\big(h(\bm x)\big).
\end{equation}
Equivalently, the set of CBF-admissible controls
\begin{equation}
\label{cbf2}
\begin{aligned}
    \mathcal U_h(\bm x)
    \triangleq
    \Big\{
    \bm u\in U
    \ \Big|\
    &\mathcal L_fh(\bm x)
    +
    \mathcal L_gh(\bm x)\bm u
    \\
    &+
    \alpha\big(h(\bm x)\big)
    \geq0
    \Big\}
\end{aligned}
\end{equation}
is nonempty for all $\bm x\in D$.
\end{definition}

The relative degree of $h(\bm x)$ with respect to the control input $\bm u$ is
the number of times it must be differentiated along the system dynamics before
$\bm u$ explicitly appears. For a function $h(\bm x)$ with relative degree
$\delta$, its higher-order Lie derivatives are given by
\begin{equation}
    \mathcal L_f^\delta h(\bm x)
    =
    \frac{d^\delta}{dt^\delta}h(\bm x),
\end{equation}
where $\mathcal L_f^\delta h(\bm x)$ denotes the $\delta$th Lie derivative of
$h$ along $\bm f$.

\begin{definition}[Exponential Control Barrier Function \cite{nguyen2016ecbf}]
A continuously differentiable function
$h:D\subset\mathbb{R}^n\to\mathbb{R}$ is an exponential control barrier
function (ECBF) if there exists a coefficient gain vector
$\bm\kappa\in\mathbb{R}^\delta$ such that, for all $\bm x\in C$,
\begin{equation}
\label{eCBF-constraint}
\begin{aligned}
    \sup_{\bm u\in U}
    \Big\{
    \mathcal L_f^\delta h(\bm x)
    &+
    \mathcal L_g\mathcal L_f^{\delta-1}h(\bm x)\bm u
    \\
    &+
    \bm\kappa\bm\eta_h(\bm x)
    \Big\}
    \geq0,
\end{aligned}
\end{equation}
where
$\bm\eta_h(\bm x)
=
[h(\bm x),\mathcal L_fh(\bm x),\ldots,
\mathcal L_f^{\delta-1}h(\bm x)]^\top$
is the Lie-derivative vector of $h(\bm x)$ and
$\bm\kappa=[k_0,k_1,\ldots,k_{\delta-1}]$.
\end{definition}

\section{Methodology}
\label{methodology}

The proposed architecture separates the docking objective from the LoS safety
requirement while retaining both within a single predictive optimization
framework. The translational tracking objective is first constructed from a
backstepping Lyapunov function so that position convergence and velocity
matching to the moving docking port are treated simultaneously. The rotating
LoS geometry is subsequently represented by two time-varying
relative-degree-two barrier functions and embedded into the MPC through ECBF
inequalities. Bounded target-attitude uncertainty is addressed directly at the
geometric level by contracting the admissible LoS corridor.

\subsection{Model-Predictive Control for Rendezvous and Docking}
\label{mpcMethodology}

The docking objective requires both the relative displacement from the port
and the relative velocity at contact to converge to zero. Let the primary
tracking error be
$\bm e_r\triangleq\bm r-\bm r_p$.
Consider first the positive-definite Lyapunov function
\begin{equation}
\label{trackingV1}
    V_1(\bm e_r)
    =
    \frac{1}{2}\bm e_r^\top\bm e_r.
\end{equation}

Since
$\dot{\bm e}_r=\bm v-\bm v_p$, the virtual velocity command is selected
according to the backstepping construction as
$\bm v_d\triangleq\bm v_p-\lambda\bm e_r$, with $\lambda>0$.
Defining the secondary error
$\bm e_v\triangleq\bm v-\bm v_d
=(\bm v-\bm v_p)+\lambda\bm e_r$
gives
\begin{equation}
\label{trackingV1dot}
    \dot V_1
    =
    -\lambda\|\bm e_r\|_2^2
    +
    \bm e_r^\top\bm e_v.
\end{equation}
Hence, $\bm e_v=\bm0$ recovers the stabilizing virtual dynamics
$\dot{\bm e}_r=-\lambda\bm e_r$, whereas regulation of $\bm e_v$
simultaneously enforces the required velocity matching.

The corresponding composite backstepping Lyapunov function is selected as
\begin{equation}
\label{trackingLyapunov}
    V_B(\bm e_r,\bm e_v)
    =
    \frac{1}{2}\bm e_r^\top\bm e_r
    +
    \frac{1}{2}\bm e_v^\top\bm e_v.
\end{equation}

Rather than deriving and applying an analytical backstepping control law, the
error structure induced by~\eqref{trackingLyapunov} is employed directly as
the tracking objective of the MPC. In particular, defining
$\bm X_p=[\bm r_p^\top,\bm v_p^\top]^\top$,
\begin{equation}
\label{backsteppingOutput}
    \bm z
    \triangleq
    \begin{bmatrix}
        \bm e_r\\
        \bm e_v
    \end{bmatrix}
    =
    \underbrace{
    \begin{bmatrix}
        \bm I_2 & \bm0_{2\times2}\\
        \lambda\bm I_2 & \bm I_2
    \end{bmatrix}}_{\bm C_\lambda}
    (\bm X-\bm X_p).
\end{equation}
Thus, the MPC penalizes the complete backstepping error vector rather than
independent position and velocity errors.

Let $T_s$ denote the sampling period and let
$\bm A_d=e^{\bm A T_s}$ and
$\bm B_d=\int_0^{T_s}e^{\bm A\tau}\bm B\,d\tau$
denote the exact zero-order-hold discretization of~\eqref{lvlhDynamics}.
For a prediction horizon $N$, the receding-horizon problem is formulated as
\begin{equation}
\label{mpcProblem}
\begin{aligned}
    \min_{\{\bm u_{j|k}\}_{j=0}^{N-1}}
    \quad&
    \sum_{j=1}^{N-1}
    \|\bm z_{j|k}\|_{\bm Q}^{2}
    +
    \sum_{j=0}^{N-1}
    \|\bm u_{j|k}\|_{\bm R}^{2}
    \\
    &+
    \|\bm z_{N|k}\|_{\bm P}^{2}
    \\
    \mathrm{s.t.}\quad&
    \bm X_{j+1|k}
    =
    \bm A_d\bm X_{j|k}
    +
    \bm B_d\bm u_{j|k},
    \\
    &
    \|\bm u_{j|k}\|_\infty
    \leq
    u_{\max},
    \qquad
    j=0,\ldots,N-1,
\end{aligned}
\end{equation}
where $\bm Q\succ0$, $\bm R\succ0$, and $\bm P\succ0$ denote the stage, input,
and terminal weighting matrices, respectively. The quadratic input penalty is
introduced to reduce unnecessary thrust expenditure while balancing the
tracking performance, whereas the LoS inequalities derived below are appended
to~\eqref{mpcProblem} along the predicted trajectory.

\subsection{Line-of-Sight Constraint}
\label{losConstraint}

\begin{figure}[htb!]
    \centering
    \includegraphics[width=\columnwidth]{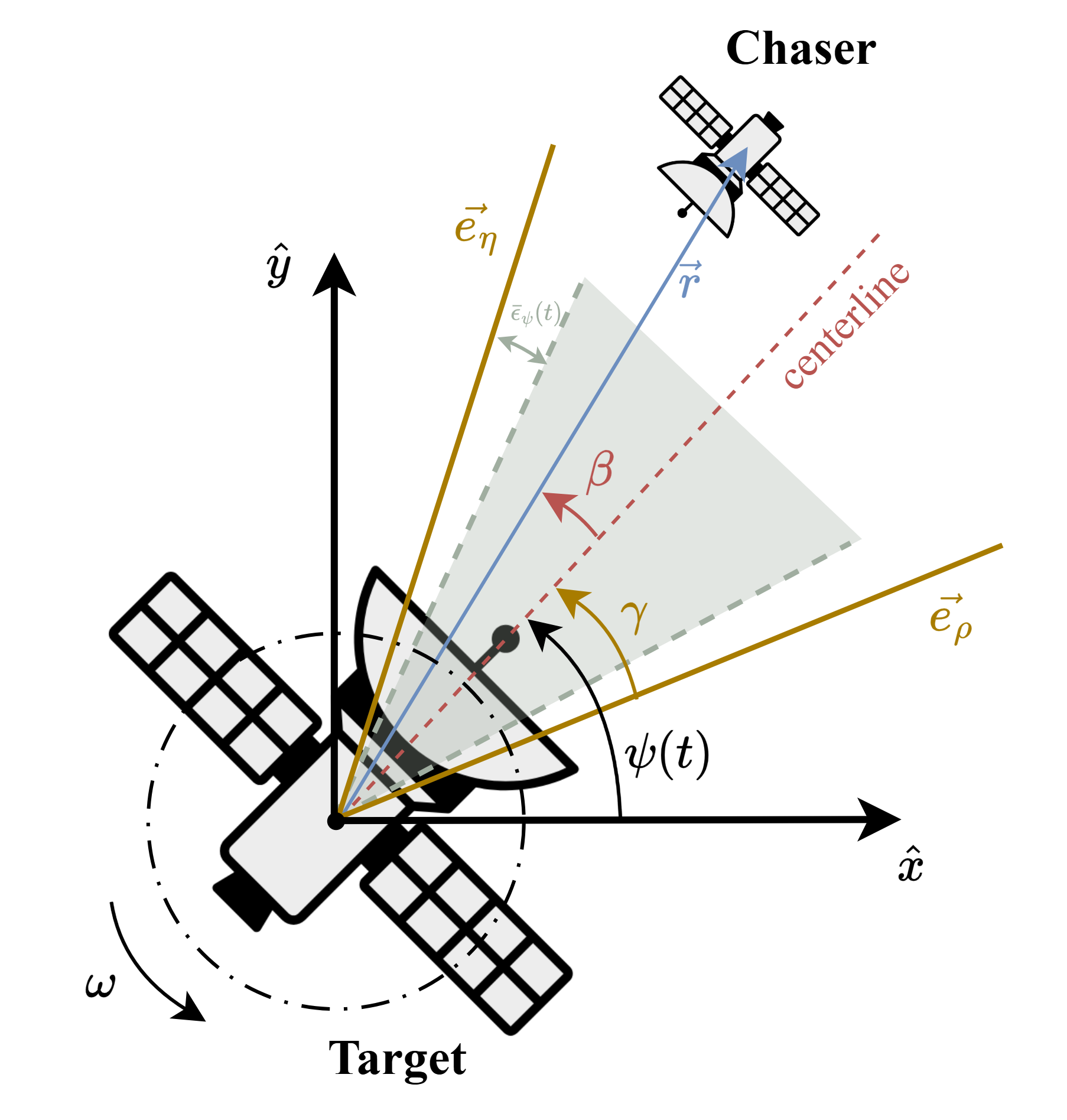}
    \caption{Geometric representation of the rotating line-of-sight
    constraint. The solid ochre rays $\bm e_\rho$ and $\bm e_\eta$
    define the nominal LoS boundaries with half-angle $\gamma$ about
    the instantaneous corridor centerline, shown by the red dashed ray.
    The target orientation relative to the LVLH $\hat{x}$ axis is
    described by $\psi(t)$ and evolves with angular rate $\omega$.
    The blue vector $\bm r$ denotes the relative position of the chaser,
    and $\beta$ is its bearing error with respect to the corridor
    centerline. The shaded region, bounded by the green dashed
    rays, represents the uncertainty-contracted admissible corridor.
    The contraction by $\bar{\epsilon}_{\psi}(t)$ accounts for the
    bounded target-attitude uncertainty, reducing the effective
    half-angle from $\gamma$ to
    $\gamma-\bar{\epsilon}_{\psi}(t)$.}
    \label{fig:los_geometry}
\end{figure}

The LoS requirement is defined as a rotating conic corridor centered on the
instantaneous docking-port bearing $\psi(t)$, as illustrated in
Fig.~\ref{fig:los_geometry}. Its apex is placed at the target center and its
half-angle is denoted by $\gamma\in(0,\pi/2)$. The two nominal boundary
directions are
$\bm e_\rho(t)=\bm e(\psi(t)-\gamma)$ and
$\bm e_\eta(t)=\bm e(\psi(t)+\gamma)$.

For two planar vectors
$\bm a,\bm b\in\mathbb{R}^2$, their oriented cross product satisfies
$\bm a\times\bm b=a_xb_y-a_yb_x=\bm b^\top\bm S\bm a$.
Accordingly, the primitive barrier functions associated with the lower and
upper LoS boundaries are
\begin{equation}
\label{primitiveLoSBarriers}
\begin{split}
    h_\rho(\bm X,t)
    &\triangleq
    \bm e_\rho(t)\times\bm r
    =
    \big(\bm S\bm e_\rho(t)\big)^\top\bm r,
    \\
    h_\eta(\bm X,t)
    &\triangleq
    \bm r\times\bm e_\eta(t)
    =
    -\big(\bm S\bm e_\eta(t)\big)^\top\bm r.
\end{split}
\end{equation}

The corresponding individual safe sets are
$C_i(t)\triangleq
\{\bm X\in\mathbb{R}^4\mid h_i(\bm X,t)\geq0\}$
for $i\in\{\rho,\eta\}$, and the complete LoS-safe set is their intersection
\begin{equation}
\label{LoSSafeSet}
    C_{\mathrm{LoS}}(t)
    \triangleq
    C_\rho(t)\cap C_\eta(t).
\end{equation}

To verify that~\eqref{LoSSafeSet} represents exactly the intended geometric
corridor, write the relative position as
$\bm r=d\,\bm e(\psi+\beta)$, where $d=\|\bm r\|_2>0$ and $\beta$ is the
bearing error measured from the corridor centerline. Substitution into
~\eqref{primitiveLoSBarriers} gives
\begin{equation}
\label{LoSBarrierAngular}
    h_\rho
    =
    d\sin(\beta+\gamma),
    \qquad
    h_\eta
    =
    d\sin(\gamma-\beta).
\end{equation}
Hence, $h_\rho\geq0$ and $h_\eta\geq0$ are jointly equivalent to
$|\beta|\leq\gamma$, and therefore the intersection in~\eqref{LoSSafeSet}
coincides with the prescribed LoS wedge.

For a unified representation, let
$\sigma_\rho=1$, $\sigma_\eta=-1$, and define
$\bm w_i\triangleq\sigma_i\bm S\bm e_i$ such that
$h_i=\bm w_i^\top\bm r$. Since $\dot\psi=\omega$, the rotating boundary
vectors satisfy
$\dot{\bm e}_i=\omega\bm S\bm e_i$ and
$\dot{\bm w}_i=-\sigma_i\omega\bm e_i$.
The first two derivatives of either barrier are therefore
\begin{equation}
\label{LoSDerivatives}
\begin{split}
    \dot h_i
    &=
    \bm w_i^\top\bm v
    -
    \sigma_i\omega\bm e_i^\top\bm r,
    \\
    \ddot h_i
    &=
    \bm w_i^\top\dot{\bm v}
    -
    2\sigma_i\omega\bm e_i^\top\bm v
    -
    \omega^2h_i.
\end{split}
\end{equation}
The control input is absent from both $h_i$ and $\dot h_i$ and enters
$\ddot h_i$ through $\dot{\bm v}$; consequently, each LoS barrier possesses
relative degree two.

Because $h_i(\bm X,t)$ is explicitly time-varying, consider the extended state
$\bar{\bm X}=[\bm X^\top,t]^\top$ with the corresponding extended vector
fields $\bar{\bm f}$ and $\bar{\bm g}$. The barrier dynamics satisfy
$\mathcal L_{\bar g}h_i=0$ and
$\mathcal L_{\bar g}\mathcal L_{\bar f}h_i=\bm w_i^\top/m_c$.
Hence, the second-order ECBF condition is
\begin{equation}
\label{LoSECBF}
    \mathcal L_{\bar f}^{2}h_i
    +
    \mathcal L_{\bar g}\mathcal L_{\bar f}h_i\,\bm u
    +
    k_1\mathcal L_{\bar f}h_i
    +
    k_0h_i
    \geq0,
\end{equation}
where $i\in\{\rho,\eta\}$ and $s^2+k_1s+k_0$ is selected to be Hurwitz.

Substitution of the LVLH acceleration in~\eqref{lvlhPartitioned} into
~\eqref{LoSDerivatives}, followed by~\eqref{LoSECBF}, yields the
control-affine LoS condition
\begin{equation}
\label{LoSAffineConstraint}
\begin{aligned}
    \frac{1}{m_c}\bm w_i^\top\bm u
    &+
    \Big[
    \bm A_r^\top\bm w_i
    +(k_0-\omega^2)\bm w_i
    -k_1\sigma_i\omega\bm e_i
    \Big]^\top\bm r
    \\
    &+
    \Big[
    \bm A_v^\top\bm w_i
    -2\sigma_i\omega\bm e_i
    +k_1\bm w_i
    \Big]^\top\bm v
    \geq0.
\end{aligned}
\end{equation}
Thus, although the LoS boundaries rotate explicitly with time, both safety
constraints remain affine in the thrust command and can be imposed directly
along the MPC prediction horizon.

\subsubsection{Attitude-Uncertainty-Aware LoS Constraint}

The target orientation is not assumed to be measured directly. Instead, the
controller propagates the nominal bearing $\psi(t)$ while the true bearing is
described by
$\psi^\star(t)=\psi(t)+\epsilon_\psi(t)$, with
$|\epsilon_\psi(t)|\leq\bar\epsilon_\psi(t)$ and
$\bar\epsilon_\psi(t)=\Delta\psi_0+\Delta\omega t<\gamma$.
The LoS region guaranteed to remain inside every corridor consistent with this
bound is
\begin{equation}
\label{robustLoSIntersection}
\begin{aligned}
    C_{\mathrm{LoS}}^r(t)
    &\triangleq
    \bigcap_{|\epsilon_\psi|
    \leq\bar\epsilon_\psi(t)}
    C_{\mathrm{LoS}}
    \big(t;\psi+\epsilon_\psi\big)
    \\
    &=
    \left\{
    \bm X\in\mathbb{R}^4
    \,\middle|\,
    |\beta|
    \leq
    \gamma-\bar\epsilon_\psi(t)
    \right\}.
\end{aligned}
\end{equation}

Hence, uncertainty is incorporated by contracting the nominal corridor rather
than by introducing an artificial disturbance into the translational
dynamics. The robust lower and upper boundary directions are
\begin{equation}
\label{robustBoundaryVectors}
    \bm e_\rho^r
    =
    \bm e\big(
    \psi-\gamma+\bar\epsilon_\psi
    \big),
    \qquad
    \bm e_\eta^r
    =
    \bm e\big(
    \psi+\gamma-\bar\epsilon_\psi
    \big).
\end{equation}

Their angular rates are
$\nu_\rho=\omega+\Delta\omega$ and
$\nu_\eta=\omega-\Delta\omega$, respectively. Defining
$\bm w_i^r=\sigma_i\bm S\bm e_i^r$, the robust barrier derivatives retain
exactly the structure in~\eqref{LoSDerivatives} after replacing
$(\bm e_i,\bm w_i,\omega)$ by
$(\bm e_i^r,\bm w_i^r,\nu_i)$. Consequently, the robust ECBF constraint
follows directly from~\eqref{LoSAffineConstraint} as
\begin{equation}
\label{robustLoSAffineConstraint}
\begin{aligned}
    \frac{1}{m_c}(\bm w_i^r)^\top\bm u
    &+
    \Big[
    \bm A_r^\top\bm w_i^r
    +(k_0-\nu_i^2)\bm w_i^r
    \Big]^\top\bm r
    \\
    &-
    k_1\sigma_i\nu_i
    (\bm e_i^r)^\top\bm r +
    \Big[
    \bm A_v^\top\bm w_i^r
    +k_1\bm w_i^r
    \Big]^\top\bm v
    \\
    &-
    2\sigma_i\nu_i
    (\bm e_i^r)^\top\bm v
    \geq0,
\end{aligned}
\end{equation}
for $i\in\{\rho,\eta\}$.
Accordingly, attitude uncertainty modifies only the geometry and time
variation of the LoS safety constraints; the CW prediction model and the
backstepping tracking objective remain unchanged.

\subsection{Overall Control Architecture}

Let
$\mathcal B_i^r(\bm X,\bm u,t)$ denote the left-hand side of the robust ECBF
inequality in~\eqref{robustLoSAffineConstraint} for
$i\in\{\rho,\eta\}$. At each sampling instant $t_k$, the moving docking-port
reference and the uncertainty-aware LoS geometry are propagated over the
prediction horizon, and the complete control problem is formulated as
\begin{equation}
\label{overallMPCBFC}
\begin{aligned}
\min_{\{\bm u_{j|k}\}_{j=0}^{N-1}}
\quad&
\sum_{j=1}^{N-1}
\|\bm z_{j|k}\|_{\bm Q}^{2}
+
\sum_{j=0}^{N-1}
\|\bm u_{j|k}\|_{\bm R}^{2}
\\
&+
\|\bm z_{N|k}\|_{\bm P}^{2}
\\
\mathrm{s.t.}\quad&
\bm X_{j+1|k}
=
\bm A_d\bm X_{j|k}
+
\bm B_d\bm u_{j|k},
\\
&
\|\bm u_{j|k}\|_\infty
\leq
u_{\max},
\\
&
\mathcal B_i^r
\big(
\bm X_{j|k},
\bm u_{j|k},
t_k+jT_s
\big)
\geq0,
\\
&
i\in\{\rho,\eta\},
\qquad
j=0,\ldots,N-1.
\end{aligned}
\end{equation}

Only the first element of the optimized control sequence is applied to the
chaser, after which the prediction horizon is shifted and
~\eqref{overallMPCBFC} is solved again using the updated state. In this
manner, the controller anticipates the moving docking reference together with
the future rotation and uncertainty contraction of the LoS corridor, while
the backstepping objective simultaneously regulates the chaser toward the
docking condition.

\section{Results}
\label{results}

The proposed architecture is evaluated in a planar rendezvous and docking
scenario using the nonlinear relative dynamics in~\eqref{nonlinearLVLH},
while the CW model in~\eqref{lvlhDynamics} is retained internally by the
predictive controller. The chaser is initialized at a range of $100$~m on the
initial docking-port bearing with zero relative velocity, whereas the target
rotates at a constant rate of $0.6^\circ$/s. The maneuver is simulated for
$400$~s, during which the propagated attitude uncertainty increases from
$1^\circ$ to $5^\circ$ under the adopted linear uncertainty model. The
principal simulation and controller parameters are summarized in
Table~\ref{tab:simulationParameters}.

\begin{table}[htb!]
\centering
\caption{Simulation and controller parameters.}
\label{tab:simulationParameters}
\small
\begin{tabularx}{\columnwidth}{@{}l X r@{}}
\hline
\textbf{Parameter} & \textbf{Description} & \textbf{Value} \\
\hline
$r_0$ & Reference-orbit radius & $6.771\times10^6$ m \\
$m_c$ & Chaser mass & $1000$ kg \\
$u_{\max}$ & Per-axis thrust limit & $250$ N \\
$\rho_p$ & Docking-port radius & $2.4$ m \\
$\omega$ & Target angular rate & $0.6^\circ$/s \\
$\psi_0$ & Initial port bearing & $180^\circ$ \\
$\gamma$ & LoS half-angle & $30^\circ$ \\
$\Delta\psi_0$ & Initial attitude-error bound & $1^\circ$ \\
$\Delta\omega$ & Angular-rate-error bound & $0.01^\circ$/s \\
$k_0$ & ECBF position gain & $2.5\times10^{-3}$ s$^{-2}$ \\
$k_1$ & ECBF rate gain & $0.1$ s$^{-1}$ \\
$\lambda$ & Backstepping rate & $0.02$ s$^{-1}$ \\
$T_s$ & MPC sampling period & $1$ s \\
$N$ & Prediction horizon & $40$ \\
$r_u$ & Input-weight scaling & $30$ \\
$t_f$ & Simulation duration & $400$ s \\
\hline
\end{tabularx}
\end{table}

\begin{figure*}[htb!]
    \centering
    \includegraphics[width=\textwidth]{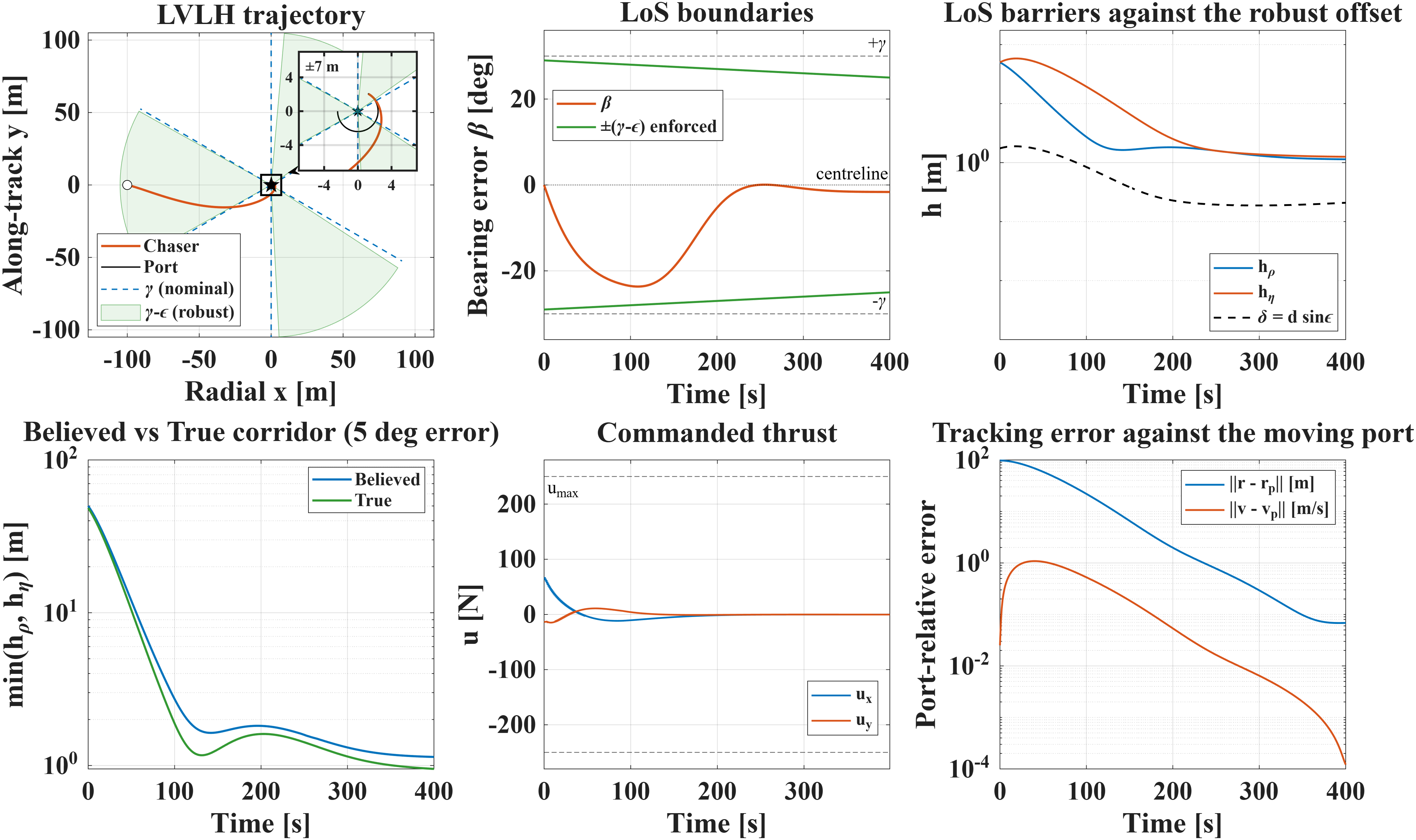}
    \caption{Closed-loop performance of the proposed robust CBF-embedded MPC
    for the rotating-port docking scenario under bounded target-attitude
    uncertainty. \textbf{Top left:} chaser trajectory in the LVLH plane. The
    dashed blue rays indicate the nominal LoS boundaries, while the shaded
    green sectors denote the uncertainty-contracted corridor enforced by the
    robust ECBF; the inset magnifies the terminal docking phase near the
    target. \textbf{Top middle:} bearing error $\beta$ measured with respect
    to the believed corridor centerline together with the enforced
    time-varying limits $\pm(\gamma-\epsilon)$; the grey dashed lines mark the
    nominal limits $\pm\gamma$. The trajectory remains strictly inside the
    contracted corridor throughout the maneuver. \textbf{Top right:} believed
    barrier values $h_\rho$ and $h_\eta$ together with the robust offset
    $\delta=d\sin\epsilon$. The inequalities $h_\rho\geq\delta$ and
    $h_\eta\geq\delta$ are maintained, which is equivalent to enforcing the
    contracted LoS geometry. \textbf{Bottom left:} comparison between the
    believed and true minimum barrier histories for the same trajectory when
    the terminal attitude mismatch reaches $5^\circ$. The gap between the
    two curves reflects the orientation-knowledge error, whereas positivity of
    the true minimum barrier confirms preservation of the actual LoS safety
    constraint. \textbf{Bottom middle:} commanded thrust components $u_x$ and
    $u_y$, both remaining within the actuator bounds $\pm u_{\max}$.
    \textbf{Bottom right:} position and velocity tracking errors relative to
    the moving docking port, showing simultaneous convergence of the chaser
    translational state to the desired docking condition.}
    \label{fig:closedLoopResults}
\end{figure*}

The closed-loop results in Fig.~\ref{fig:closedLoopResults} are interpreted
from three complementary perspectives. The trajectory, bearing-error, and
barrier histories first indicate whether the predictive controller preserves
the uncertainty-contracted LoS corridor throughout the approach. The believed
and true barrier histories then show whether the imposed geometric contraction
is sufficient to accommodate the accumulated attitude uncertainty. Finally,
the commanded thrust and moving-port tracking errors characterize the control
effort required by the maneuver and the resulting terminal tracking
performance.

Figure~\ref{fig:closedLoopResults} shows that the chaser remains inside the
uncertainty-contracted corridor throughout the maneuver while converging
toward the moving docking port. The bearing-error and barrier histories show
that the propagated attitude uncertainty is accommodated without violation of
the true LoS constraint, whereas the commanded thrust remains within the
available control authority. Simultaneous convergence of the position and
velocity errors further demonstrates successful tracking of the rotating
docking reference.

The fuel expenditure is evaluated through the total absolute impulse
$J_{\mathrm{imp}}=\int_0^{t_f}(|u_x(t)|+|u_y(t)|)\,dt$, which is proportional
to the consumed propellant for a fixed specific impulse under the adopted
axis-wise thrust model. For the selected input weighting $r_u=30$, the
component-wise impulses are $J_x=2197.68$~N\,s and $J_y=1003.24$~N\,s,
corresponding to $J_{\mathrm{imp}}=3200.91$~N\,s. Under the same tracking and
safety formulation, reducing the input weighting to $r_u=3$ increases the
total impulse to $4247.09$~N\,s. The adopted control-effort penalization
therefore reduces the total impulse by $24.63\%$, while preserving LoS safety
and a terminal position error of $0.0677$~m.

\section{Conclusion \& Future Work}

This study addressed autonomous rendezvous and docking with a rotating target,
where the chaser must track a moving docking port while satisfying
time-varying LoS and actuator constraints under imperfect target-orientation
knowledge. A CBF-embedded MPC architecture was developed in which a
backstepping-based objective regulates the position and velocity errors
relative to the moving port, while relative-degree-two ECBF constraints enforce
the rotating LoS corridor over the prediction horizon. The effect of bounded
attitude and angular-rate uncertainty was incorporated directly into the
safety geometry through a time-varying contraction of the admissible corridor,
without modifying the nominal prediction dynamics or tracking objective.

Closed-loop simulations using the nonlinear relative dynamics showed that the
chaser remained inside the uncertainty-contracted corridor as the propagated
orientation uncertainty increased from $1^\circ$ to $5^\circ$, while the
thrust commands remained within the prescribed $250$~N per-axis limit and the
terminal position error was reduced to $0.0677$~m. The control-effort penalty
also provided a measurable reduction in thrust expenditure: the selected
weighting $r_u=30$ required a total absolute impulse of
$3200.91$~N\,s, compared with $4247.09$~N\,s for $r_u=3$, corresponding to a
$24.63\%$ reduction while preserving the LoS safety constraint. These results
demonstrate that the proposed formulation can simultaneously accommodate the
moving docking objective, uncertainty-aware safety requirements, bounded
actuation, and reduced thrust expenditure within a single predictive control
framework.

Future work will extend the formulation to full three-dimensional relative
and attitude dynamics, including targets undergoing multi-axis tumbling, as
encountered in active debris-removal scenarios involving non-cooperative and
freely rotating objects. Camera-based relative navigation will also be
incorporated for terminal docking, while explicit spacecraft-body geometry
will be introduced to impose collision-avoidance constraints in addition to
the LoS requirements.

\bibliography{references}

@inproceedings{park2011mpc,
  author    = {Park, Hyeongjun and Di Cairano, Stefano and Kolmanovsky, Ilya},
  title     = {Model Predictive Control for Spacecraft Rendezvous and Docking with a Rotating/Tumbling Platform and for Debris Avoidance},
  booktitle = {2011 American Control Conference (ACC)},
  address   = {San Francisco, CA, USA},
  pages     = {1922--1927},
  year      = {2011},
  doi       = {10.1109/ACC.2011.5991151}
}

@article{bashnick2023fastmpc,
  author  = {Bashnick, C. and Ulrich, S.},
  title   = {Fast Model Predictive Control for Spacecraft Rendezvous and Docking with Obstacle Avoidance},
  journal = {Journal of Guidance, Control, and Dynamics},
  volume  = {46},
  number  = {5},
  pages   = {998--1007},
  year    = {2023},
  doi     = {10.2514/1.G007314}
}

@article{fear2024mpc,
  author  = {Fear, A. and Lightsey, E. G.},
  title   = {Autonomous Rendezvous and Docking Implementation for Small Satellites Using Model Predictive Control},
  journal = {Journal of Guidance, Control, and Dynamics},
  volume  = {47},
  number  = {3},
  pages   = {539--547},
  year    = {2024},
  doi     = {10.2514/1.G007523}
}

@article{specht2023tubempc,
  author  = {Specht, Caroline and Bishnoi, Abhiraj and Lampariello, Roberto},
  title   = {Autonomous Spacecraft Rendezvous Using Tube-Based Model Predictive Control: Design and Application},
  journal = {Journal of Guidance, Control, and Dynamics},
  volume  = {46},
  number  = {7},
  pages   = {1243--1261},
  year    = {2023},
  doi     = {10.2514/1.G007280}
}

@article{albee2021robust,
  author  = {Albee, Keenan and Oestreich, Christopher and Specht, Christopher and Ter{\'a}n Espinoza, Alejandro and Todd, John and Hokaj, Ilir and Lampariello, Roberto and Linares, Richard},
  title   = {A Robust Observation, Planning, and Control Pipeline for Autonomous Rendezvous with Tumbling Targets},
  journal = {Frontiers in Robotics and AI},
  volume  = {8},
  pages   = {641338},
  year    = {2021},
  doi     = {10.3389/frobt.2021.641338}
}

@article{dong2022safempc,
  author  = {Dong, K. and Luo, J. and Lim{\'o}n, D.},
  title   = {A Novel Stable and Safe Model Predictive Control Framework for Autonomous Rendezvous and Docking with a Tumbling Target},
  journal = {Acta Astronautica},
  volume  = {200},
  pages   = {176--187},
  year    = {2022},
  doi     = {10.1016/j.actaastro.2022.08.012}
}

@inproceedings{agrawal2021iccbf,
  author    = {Agrawal, Devansh R. and Panagou, Dimitra},
  title     = {Safe Control Synthesis via Input Constrained Control Barrier Functions},
  booktitle = {2021 60th IEEE Conference on Decision and Control (CDC)},
  address   = {Austin, TX, USA},
  pages     = {6113--6118},
  year      = {2021},
  doi       = {10.1109/CDC45484.2021.9682938}
}

@article{vinod2018tradeoff,
  author  = {Vinod, Abraham P. and Oishi, Meeko M. K.},
  title   = {Optimal Trade-off Analysis for Efficiency and Safety in the Spacecraft Rendezvous and Docking Problem},
  journal = {IFAC-PapersOnLine},
  volume  = {51},
  number  = {12},
  pages   = {136--141},
  year    = {2018},
  doi     = {10.1016/j.ifacol.2018.07.101}
}

@article{zappulla2019aapf,
  author  = {Zappulla, Richard II and Park, Hyeongjun and Virgili-Llop, Josep and Romano, Marcello},
  title   = {Real-Time Autonomous Spacecraft Proximity Maneuvers and Docking Using an Adaptive Artificial Potential Field Approach},
  journal = {IEEE Transactions on Control Systems Technology},
  volume  = {27},
  number  = {6},
  pages   = {2598--2605},
  year    = {2019},
  doi     = {10.1109/TCST.2018.2866963}
}

@phdthesis{sternberg2017docking,
  author  = {Sternberg, David Charles},
  title   = {Optimal Docking to Tumbling Objects with Uncertain Properties},
  school  = {Massachusetts Institute of Technology},
  address = {Cambridge, MA, USA},
  year    = {2017}
}

@article{ames2017cbf,
  author  = {Ames, Aaron D. and Xu, Xiangru and Grizzle, Jessy W. and Tabuada, Paulo},
  title   = {Control Barrier Function Based Quadratic Programs for Safety Critical Systems},
  journal = {IEEE Transactions on Automatic Control},
  volume  = {62},
  number  = {8},
  pages   = {3861--3876},
  year    = {2017},
  doi     = {10.1109/TAC.2016.2638961}
}

@inproceedings{nguyen2016ecbf,
  author    = {Nguyen, Quan and Sreenath, Koushil},
  title     = {Exponential Control Barrier Functions for Enforcing High Relative-Degree Safety-Critical Constraints},
  booktitle = {2016 American Control Conference (ACC)},
  address   = {Boston, MA, USA},
  pages     = {322--328},
  year      = {2016},
  doi       = {10.1109/ACC.2016.7524935}
}

@article{altunkaya2025nuisance,
  author  = {Altunkaya, Ege C. and {\"O}zkol, {\.I}brahim},
  title   = {Nuisance-Free Ground Collision Avoidance Design with Adaptive
             Exponential Control Barrier Functions},
  journal = {Journal of Guidance, Control, and Dynamics},
  volume  = {48},
  number  = {12},
  pages   = {2693--2707},
  year    = {2025},
  doi     = {10.2514/1.G009218}
}

\end{document}